\documentclass[preprint,nofootinbib]{revtex4-1}
\usepackage{amssymb,amsfonts}
\usepackage{graphicx}

\newcommand{\be}{\begin{equation}}
\newcommand{\ee}{\end{equation}}

\newcommand{\cl}{\mathcal{L}}

\newcommand{\mr}{\mathbb{R}}

\newcommand{\ch}{\mathcal{H}}

\newcommand{\bi}{\begin{itemize}}
\newcommand{\ei}{\end{itemize}}

\newcommand{\bnum}{\begin{enumerate}}
\newcommand{\enum}{\end{enumerate}}

\newcommand{\ba}{\begin{eqnarray}}
\newcommand{\ea}{\end{eqnarray}}

\newcommand{\eq}[1]{(\ref{#1})}

\begin{document}

\title{Holographic entanglement entropy and thermodynamic instability of planar R-charged black holes}

\author{Xing Wu}
\email{xwu@itp.ac.cn}
\affiliation{ Key Laboratory of Frontiers in Theoretical Physics, Institute of Theoretical Physics, Chinese Academy of Sciences, P.O. Box 2735, Beijing 100190, China.   \\ }

\begin{abstract}The holographic entanglement entropy of an infinite strip subsystem on the asymptotic AdS boundary is used as a probe to study the thermodynamic instabilities of planar R-charged black holes (or their dual field theories). We focus on the single-charge AdS black holes in $D=5$, which correspond to spinning D3-branes with one non-vanishing angular momentum. Our results show that the holographic entanglement entropy indeed exhibits the thermodynamic instability associated with the divergence of the specific heat. When the width of the strip is large enough, the finite part of the holographic entanglement  entropy as a function of the temperature resembles the thermal entropy, as is expected. As the width becomes smaller, however, the two entropies behave differently. In particular, there exists a critical value for the width of the strip, below which the finite part of the holographic entanglement entropy as a function of the temperature develops a self-intersection. We also find similar behavior in the single-charge black holes in $D=4$ and $7$.
\end{abstract}

\maketitle

\section{Introduction}

Within the frame work of the AdS/CFT correspondence \cite{Maldacena:1997re,Witten:1998qj,Gubser:1998bc}, the holographic entanglement entropy \cite{Ryu:2006bv,Ryu:2006ef}  (c.f. \cite{Casini:2011kv,Lewkowycz:2013nqa,Caputa:2013eka} for recent works related to the proof of this proposal) has been used as a probe for systems with phase transitions, such as the holographic superconductor \cite{Albash:2012pd,Cai:2012nm} and more recently the Reissner-Nordstr\"om AdS (RNAdS) black holes \cite{Johnson:2013dka}. In these systems, a typical feature is that within a certain range of parameters, there exist more than one branch of solutions (different phases) having the same temperature. Usually, the most physically relevant solution should satisfy two criterions: (1) local stability in the sense \cite{Gubser:1998jb,Cvetic:1999ne} that the thermal entropy as a function of the extensive thermodynamic variables $x_i$ should be subadditive, in particular, when $S$ is a smooth function, the Hessian $\frac{\partial^2 S}{\partial x_i\partial x_j}$ should be negative definite; and (2) global stability in the sense that the free energy should be minimized. On the other hand, solutions failing to satisfy the local stability condition is not physically realizable, and those obeying (1) but not (2) are metastable and will decay into other configurations with lower free energy. We will focus on the local instability (1) in this paper. Thermodynamic instabilities are usually associated with the divergence of quantities like specific heat or charge susceptibility, etc. The work initiated in \cite{Albash:2012pd} illustrated how the holographic entanglement entropy can be a useful tool to exhibit various phase transitions of some field theories dual to black holes in global AdS. Since the definition of the entanglement entropy is applicable even when thermodynamical quantities are not well-defined, e.g. in processes out of equilibrium or in quantum phase transitions, this method may have potential application in more general context, for example, see  \cite{AbajoArrastia:2010yt,Albash:2010mv,Galante:2012pv,Caceres:2012em} for its application in holographic thermalization \cite{Balasubramanian:2010ce,Balasubramanian:2011ur}. In this paper, we use this method to study thermodynamic instability of the planar R-charged black holes in AdS (or their dual field theories).

The planar R-charged black hole solution in $D=5$ studied here is the STU model which was first obtained as a special case of the solution in $D=5,\mathcal{N}=2$ gauged supergravity in  \cite{Behrndt:1998jd}, where it was argued that this model is also a solution of gauged supergravity with more supersymmetries, in particular, $\mathcal{N}=8$ supersymmetries. In fact, the $D=5,\mathcal{N}=8$ gauged supergravity can be obtained from the Kaluza-Klein reduction  on $S^5$ of $D=10$ type IIB supergravity, where the isometry group of the $S^5$ precisely gives the $SO(6)$ gauge symmetry of the five dimensional theory, which further gives the R-symmetry group of the CFT via the AdS/CFT correspondence. As a consistent truncation, one can turn on only the gauge fields of the Cartan subalgebra of the full gauge group, which is $U(1)^3$ for $SO(6)$. In this way, one obtains a correspondence between  AdS$_5$ black holes with three U(1) charges (i.e. R-charges), and  spinning D3-branes with three independent angular momenta orthogonal to the branes \cite{Cvetic:1999ne,Cvetic:1999xp}. The thermodynamics of such R-charged black hole/D3-brane system has been studied in \cite{Gubser:1998jb,Cai:1998ji,Cvetic:1999ne,Cvetic:1999rb,Sahay:2010yq}, where it was found that for certain range of angular momenta/R-charges and energy densities, the system develops thermodynamic instability in the sense that the entropy fails to be subadditive, i.e. the Hessian of the entropy as a function of the thermodynamic variables has at least one positive eigenvalue.  As we shall see below, the holographic entanglement entropy can also serve as a tool to probe some (but not all, see Discussion) of these instabilities.

In fact, the above equivalence of R-charged black holes in $D=5$ and spinning D3-branes can be extended to R-charged black holes in $D=4$ ($D=7$) and spinning M2-branes (M5-branes) \cite{Cvetic:1999ne,Cvetic:1999xp}. We will also briefly discuss the holographic entanglement entropy in this context.

This paper is organized as follows. We first review the thermodynamics of the planar R-charged black holes in $D=5$ in the next section. Then in section \ref{sec 3}, we calculate the holographic entanglement entropy in $D=5$ with a single charge in the grand canonical and canonical ensembles, respectively. After extending our discussion to the context of R-charged black holes in $D=4$ and $7$, we conclude this paper with discussions in the last section.

\section{Thermodynamics of Planar R-Charged Black Holes}
Here we review some relevant properties of the thermodynamics of planar R-charged black holes. More details can be found in \cite{Cvetic:1999ne,Cvetic:1999rb,Sahay:2010yq}. The general solution with three R-charges is given by
\be
ds_5^2 = -\ch^{-2/3}\, f\, dt^2 +
\ch^{1/3}\, (f^{-1}\, d r^2 +  r^2 d\Omega^2_{3,k}),
\ee
with
\be
f=k-\frac{\mu}{ r^2} +  \frac{r^2}{L^2}\, \ch\ ,\qquad
\ch= H_1H_2H_3,\ \ \ H_i = 1 + \frac{q_i}{\,  r^2},\ \ q_i= \mu\, \sinh^2\beta_i,
\ee
where $L$ is the AdS curvature radius, which will be set to unity for convenience in the following, $\mu$ and $\beta_i$ are parameters of the solution, and $d\Omega_{3,1}^2$  gives the standard metric on $S^3$ for $k=1$, while $d\Omega_{3,0}^2=d\vec y_3^2$ on $\mr^3$ for $k=0$. Note in passing that the case with $k=-1$ is studied in \cite{Cai:2004pz}. Requiring $f(r_+)=0$ gives the horizon $r_+$.  The gauge fields associated with the $U(1)^3$ are
\be
A^i =\frac{Q_i}{r^2+q_i} dt, \ \ \ i=1,2,3.
\ee
The physical charge $Q_i$ when $k=1$ is given by
\be
Q_i=\mu \sinh\beta_i\cosh\beta_i=\sqrt{q_i(\mu+q_i)}.
\ee
As argued in \cite{Cvetic:1999ne,Witten:1998zw}, the $k=0$ case can be obtained by taking the large black hole limit in the $k=1$ case, or equivalently, taking the small $\beta_i$ limit, which gives
\be
Q_i=\sqrt{\mu q_i}.
\ee
In the following, we will only consider the planar case with $k=0$.\footnote{
There are also scalar fields in the solution. But they do not concern us since they are irrelevant for our purpose to study the holographic entanglement entropy.}

The ADM mass of the black hole is
\be
M=\frac{V_3}{8\pi G}\frac{3}{2}\mu.
\ee
The temperature of the black hole is
\be
T=\frac{r_+^2(-\frac{\prod_i\rho_i^2}{ r_+^4}+\frac{\prod_j\rho_j^2}{ r_+^2}\sum_i\frac{1}{\rho_i^2})}{2\pi \prod_i\rho_i},\ \ \ \rho_i^2\equiv r_+^2+q_i.
\ee
The thermal entropy is
\footnote{
We assume the convention $8\pi G=1$. Note that the volume $V_3$ of the $\mr^3$ is essentially divergent. Thus the physically sensible quantities should really be their densities. Bearing this in mind, we will simply set $V_3=1$ for convenience.
}
\be
S_{th}=\frac{A}{4G }=2\pi V_3 \prod_i(r_+^2+q_i)^{1/2}.
\ee
The electric potential at the horizon, which plays the role of the thermodynamic conjugate variable to the charge $Q_i$, is
\be
\phi_i\equiv A_t^i(r_+)=\frac{Q_i}{r_+^2+q_i}.
\ee

For simplicity, we will focus on the single-charge case with
$q_1=q$, $q_2=q_3=0$. Then the entropy becomes
    \be
{S_{th}}=2\pi r_+^2\sqrt{r_+^2+q}.
\ee
The physical charge is
\be\label{Case1 Q}
Q= \sqrt{(r_+^2+q)q r_+^2}.
\ee
The potential is
\be \label{Case1 phi}
\phi =\frac{\sqrt{q} r_+}{\sqrt{r_+^2+q}}.
\ee
The temperature is
\be
T(r_+,q)=\frac{1}{2\pi}\frac{2r_+^2+q}{\sqrt{r_+^2+q}}.
\ee
In the grand canonical ensemble, one should regard $q=q(\phi,r_+)$ and use (\ref{Case1 phi}) to get
\be
T(r_+,\phi)=\frac{1}{2\pi}\frac{2r_+^2-\phi^2}{\sqrt{r_+^2-\phi^2}},
\ee
where $r_+\geq \phi$ is  required to ensure a physically sensible temperature. It has a non-zero minimum $T_{min}=\phi\sqrt{2}/\pi\approx0.45 \phi$.
So there is no extremal limit in this case.
In the canonical ensemble one should regard $q=q(Q,r_+)$ and use (\ref{Case1 Q}) to get
\be
T(r_+,Q)=\frac{3r_+^3+\sqrt{r_+^6+4Q}}{2\pi\sqrt{2r_+(r_+^3+\sqrt{r_+^6+4Q})}}.
\ee
where $r_+$ starts from zero. There is also a non-vanishing minimal $T_{min}\approx0.26 Q^{1/3}$.

The specific heat with fixed $\phi$ and $Q$ are respectively given by (parameterized in terms of $q$ and $r_+$)
\be\label{C phi}
C_\phi=T\left(\frac{\partial S_{th}}{\partial T}\right)_\phi=2\pi\frac{r_+^2(q+2r_+^2)(q-3r_+^2)}{(q-2r_+^2)\sqrt{r_+^2+q}}.
\ee
\be\label{C Q}
C_Q=T\left(\frac{\partial S_{th}}{\partial T}\right)_Q=6\pi\frac{r_+^2(r_+^2+q)^{3/2}(q+2r_+^2)}{2r_+^4+5qr_+^2-q^2}.
\ee

\begin{figure}[htbp]
\begin{center}
        \includegraphics[scale=0.7]{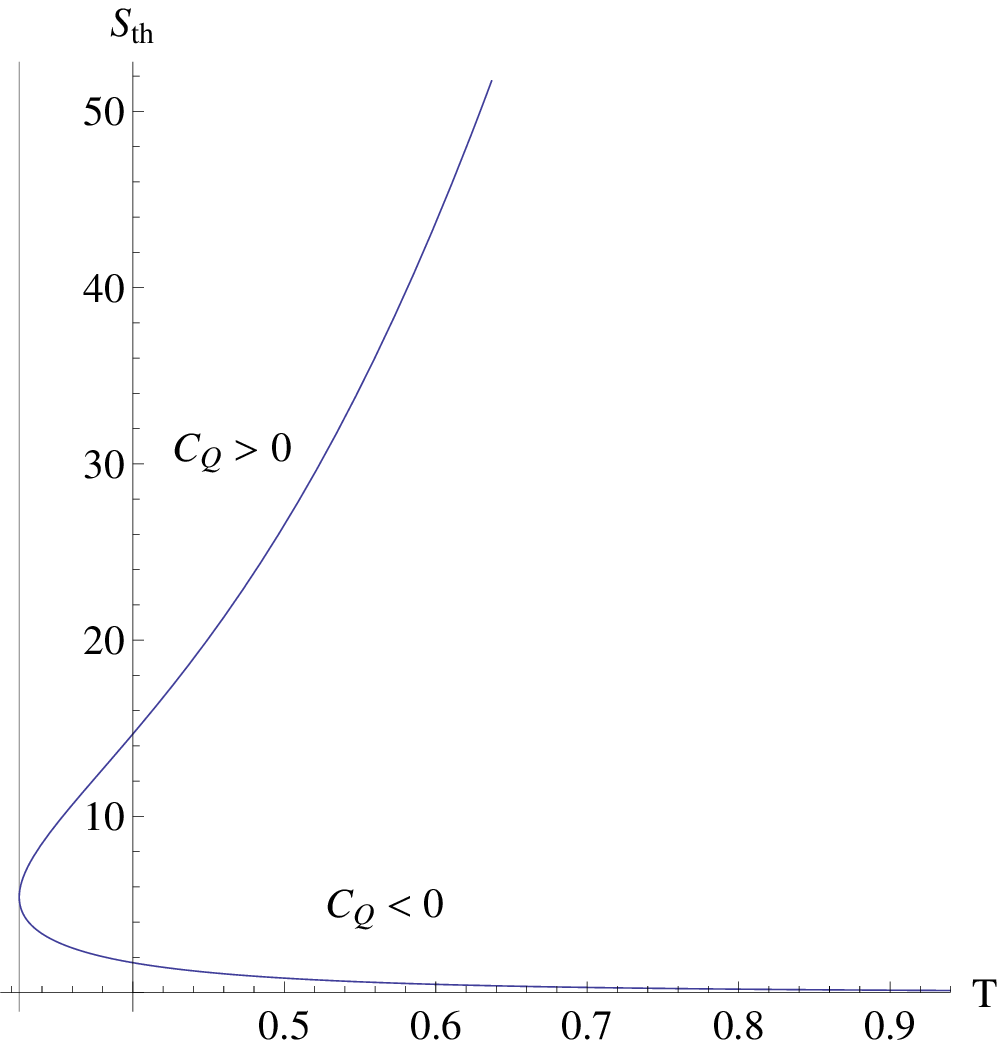}
        \includegraphics[scale=0.7]{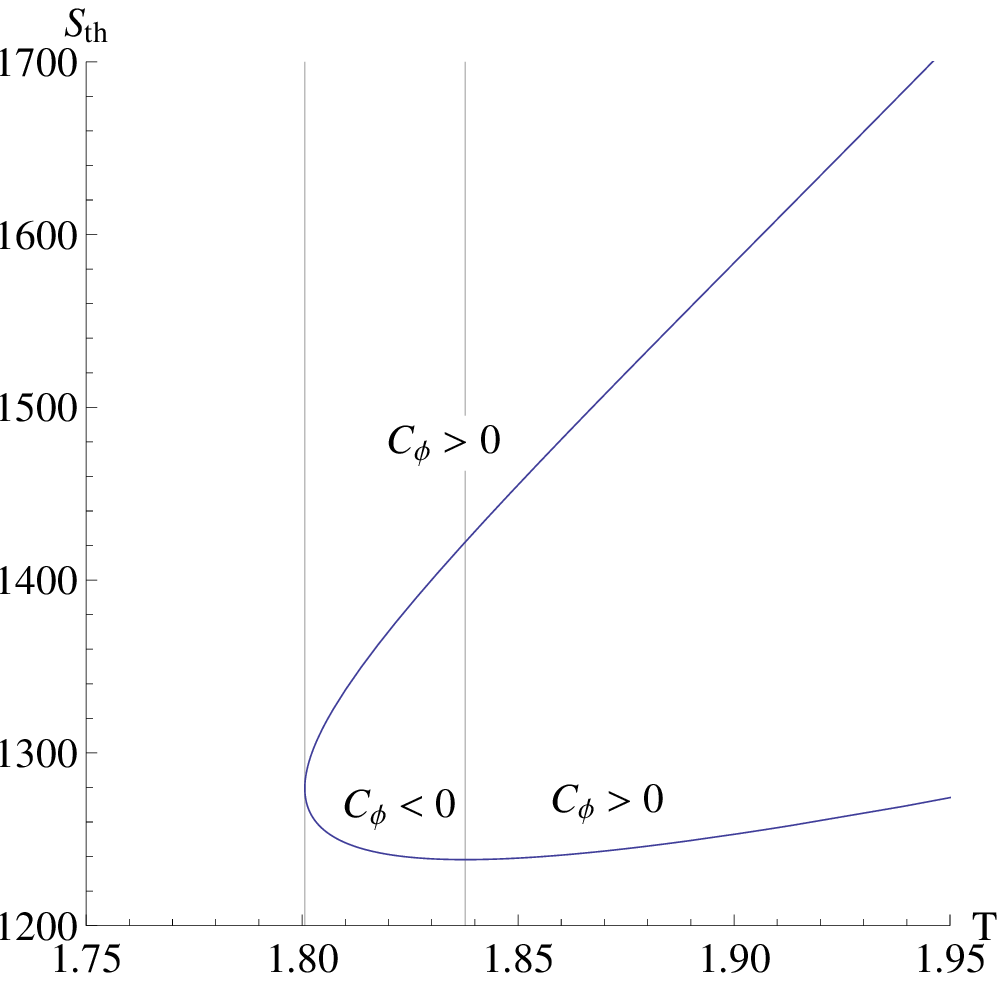}
        \caption{ Thermal entropy $S_{fin}(T)$. Left: the canonical ensemble with fixed $Q=2$. The vertical line corresponds to $T_{min}\approx0.32$. Right: the grand canonical ensemble with fixed $\phi=4$. The first vertical line corresponds to $T_{min}\approx1.8$, and the second corresponds to $T_1\approx1.84$ where the specific heat of the lower branch changes sign.  }  \label{Case1Sthermal}
\end{center}
\end{figure}

    The typical behavior of the entropy as a function of the temperature is plotted in Fig.  \ref{Case1Sthermal}. In the canonical ensemble with fixed $Q=2$, the upper branch corresponds to large black holes which are stable, while the lower branch corresponds to small black holes whose specific heat is negative and therefore are thermodynamically unstable. Moreover, the specific heat, which corresponds to the slope of the curve, diverges at the minimal temperature $T_{min}\approx 0.33$. In the grand canonical ensemble with fixed $\phi=4$, a new feature is that the lower small black hole branch increases with $T$ after $T_1\approx 1.84$. This indicates that the specific heat becomes positive for the small black holes as $T\geq T_1$. Such behavior is consistent with the result of the analysis of spinning D3-branes thermodynamics \cite{Cai:1998ji,Cvetic:1999rb}, where the parameter used there is $\ell$, i.e. the angular momentum parameter of the spinning D3-branes, which is related to our parameter here by $\ell^2=q$. In particular, using \eq{C phi}, the upper branch here corresponding to $q<2r_+^2$ is equivalent to $\ell^2<2r_+^2$ there, and the lower branch after $T\geq T_1$ corresponding to $q>3r_+^2$ is equivalent to $\ell^2>3r_+^2$ there. Although this lower branch has a positive specific heat, it is still unstable in the sense that the entropy is not subadditive \cite{Cvetic:1999rb}. Indeed, as calculated in \cite{Cai:1998ji}, the isothermal capacitance can still become negative when $\ell^2>3r_+^2$.

\begin{figure}
\begin{center}
        \centering
        \includegraphics[scale=0.7]{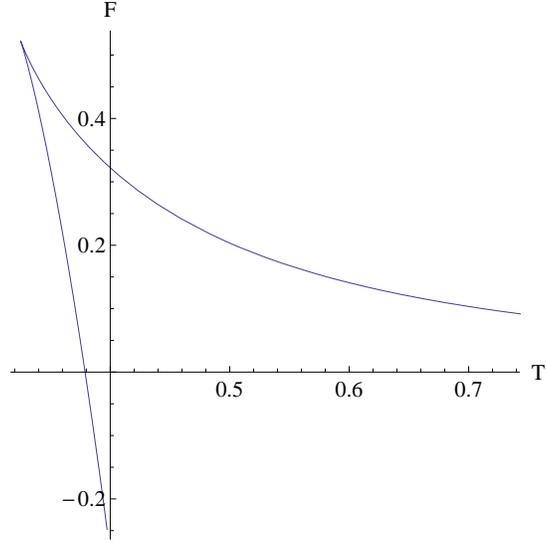}

        \caption{The Helmholtz free energy $F(T)$  with $Q=2$.}
        \label{FreeEnergyF}
\end{center}
\end{figure}

\begin{figure}
 \begin{center}
        \centering
          \includegraphics[scale=0.7]{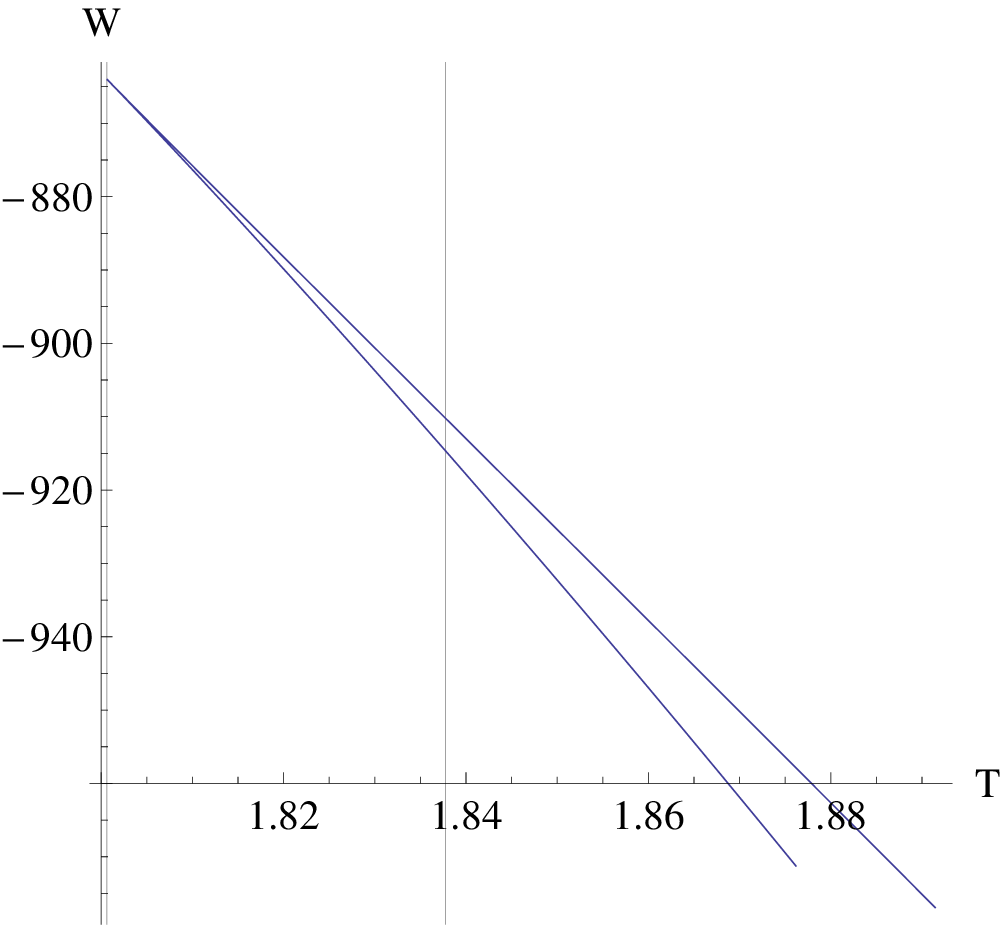}
          \includegraphics[scale=0.7]{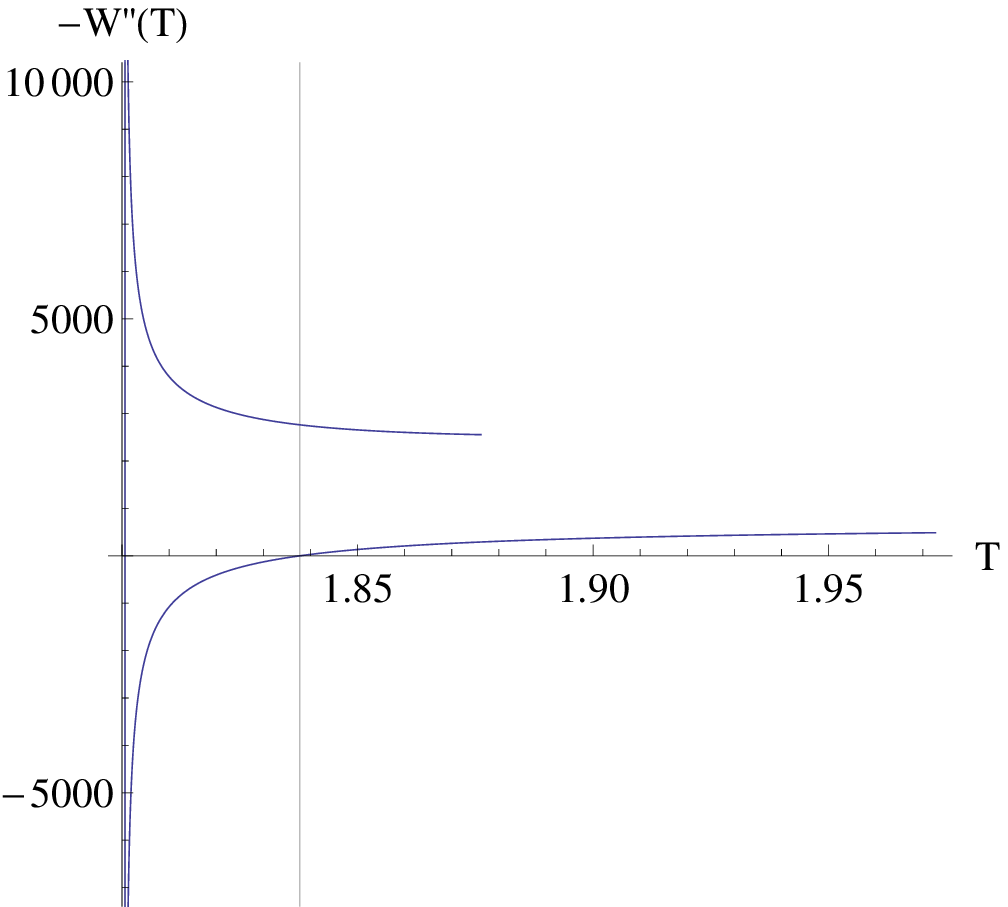}

        \caption{The Gibbs free energy $W(T)$ and its second order derivative $-W''(T)$   with $\phi=4$. The vertical line denotes the point $T_1$ where the second derivative $W''(T)$ of the upper branch changes sign. }
        \label{FreeEnergyW}
 \end{center}
 \end{figure}

        The Helmholtz free energy $F$ and the Gibbs free energy $W$ are given by \cite{Sahay:2010yq,Cai:1998ji}
\be
F=-\frac{V_3}{16\pi G}r_+^2(r_+^2-q),
\ee
\be
W=-\frac{V_3}{16\pi G}r_+^2(r_+^2+q),
\ee
which are related by a Legendre transformation $W=F-\phi Q$.
They are plotted as functions of $T$ in Fig.  \ref{FreeEnergyF} and \ref{FreeEnergyW}, where one can see clearly that the large black hole branch always has a lower free energy than the small one and therefore is globally favored. In particular, the second order derivative
\footnote
 {In Fig. \ref{FreeEnergyW} we plot $-W''(T)$ instead of $W''(T)$ so that it essentially corresponds to the specific heat $C_\phi$.
}
$W''(T)$ of the upper branch (small black holes) changes sign across the temperature $T_1$ denoted by a vertical line in Fig.  \ref{FreeEnergyW}, indicating a change of sign of the specific heat, in accordance with the behavior of $S_{th}(T)$ in Fig.  \ref{Case1Sthermal}.

\section{Holographic Entanglement Entropy}
\label{sec 3}

\subsection{Basics and Setup}
In a general quantum system described by a density matrix $\rho$ and  composed of a subsystem $A$ and its complement $B$, the entanglement entropy of $A$ is given by
\be
S_A=-Tr_A(\rho_A\ln \rho_A),
\ee
where $\rho_A=Tr_B \rho$ is the reduced density matrix, and $Tr_X$ means to trace over the subsystem $X$. For a general quantum field theory, the calculation of $S_A$ is very cumbersome. In the context of the AdS/CFT correspondence, however, the entanglement entropy of a CFT system can be calculated in an elegant way following the proposal of the holographic entanglement entropy \cite{Ryu:2006bv,Ryu:2006ef}. More precisely, for a CFT$_d$ dual to some static asymptotic AdS$_{d+1}$ spacetime with a timelike Killing field parameterized by $t$, its entanglement entropy of a subsystem $A$ within the spatial region bounded by $\partial A$ is given by
\be
S_A=\frac{{\rm Area}(\gamma)}{4 G },\ \ \ {\rm Area}(\gamma)=\int_\gamma \sqrt{h}d^{d-1}x,
\ee
where $\gamma$ is a ($d-1)$-dimensional surface anchored on the boundary with $\partial\gamma=\partial A$, which is the minimal surface in the bulk (more precisely, in the $d$-dimensional constant-$t$ hypersurface),  $h$ is the determinant of the induced metric on $\gamma$, and $G $ is the $(d+1)$-dimensional Newton's constant. Note the generalization to time-dependent cases has been studied in \cite{Hubeny:2007xt} where the minimal surface condition is replaced by an extremal surface condition.

In our case of the R-charged black holes in AdS$_5$, which is dual to some CFT$_4$ on the boundary with the R-symmetry, let the subsystem   be a strip between $x=\pm l/2$ and extended along the $\vec y_2$ directions. By symmetry, we assume that the minimal surface is given by $r=r(x)$, which satisfies the boundary conditions
\be
r(\pm l/2)=r_c,\ \ \ \dot r( 0)=0.
\ee
where $r_c$ is the  UV cutoff near the boundary $r\rightarrow\infty$. We will only consider the branch $x\in[0,l/2]$ where $\dot r\geq0$, while the other branch can be trivially obtained by symmetry.

Then the induced geometry on the surface $\gamma$ becomes
\be
ds^2_{3}=\ch^{1/3}\left[ (\frac{\dot r^2}{f}+r^2)dx^2+r^2d\vec y_{2}^2\right].
\ee
The area is given by
\be
A=2V_{2}\int^{l/2}_0dx\cl,\ \ \ \cl=r^{2}\sqrt{\ch(\frac{\dot r^2}{f}+r^2)}.
\ee
Since there is no explicit dependence of $\cl$ on $x$, we have a conserved quantity
\be
\cl-\dot r\frac{\partial \cl}{\partial \dot r}=\frac{\sqrt{\ch}r^{4}}{\sqrt{\dot r^2/f+r^2}}.
\ee
Using the above boundary conditions, we obtain the equation of motion for $r(x)$
\be
\dot r=\sqrt{\left( \frac{\ch}{\ch_*}\frac{r^{8}}{r_*^{6}}-r^2\right)f},
\ee
where $r_*$ denotes the turning point $r_*\equiv r( 0)$, and $\ch_*\equiv\ch(r_*)$.

Now the entanglement entropy $S_A$ as a function of the size $l$ is obtained as parameterized by $r_*$
\ba
l(r_*)&=&2\int^\infty_{r_*}\frac{dr}{\sqrt{\left(\frac{\ch}{\ch_*}\frac{r^{8}}{r_*^{6}}-r^2\right)f}}, \\
S_A(r_*)&=&\frac{V_{2}}{2G}\int^{r_c}_{r_*} r^{2}\sqrt{\ch}\sqrt{\frac{\ch}{\ch_*}\frac{r^{8}}{r_*^{6}}}\frac{dr}{\sqrt{\left(\frac{\ch}{\ch_*}\frac{r^{8}}{r_*^{6}}-r^2\right)f}}.
\ea
Since $V_{2}$ is essentially infinite, the physically sensible quantity should be the entanglement entropy density defined by $S_A/V_{2}$. In the following we will simply set $V_{2}=1$ and still refer to $S_A$ and $S_{fin}$ as entanglement entropy. There is a UV divergence in $S_A$    as $r_c\rightarrow\infty$. Simple analysis of the integral in $S_A$ at the large $r$ limit shows that it can be regularized as
\be
S_A(r_*)=\frac{r_c^{2}}{4G}+S_{fin}(r_*),
\ee
where $S_{fin}$ is the finite part, which is independent of the UV cutoff $r_c$.

\subsection{Single-Charge Black Holes in AdS$_5$}
\begin{figure}
\begin{center}
        \centering
        \includegraphics[scale=0.7]{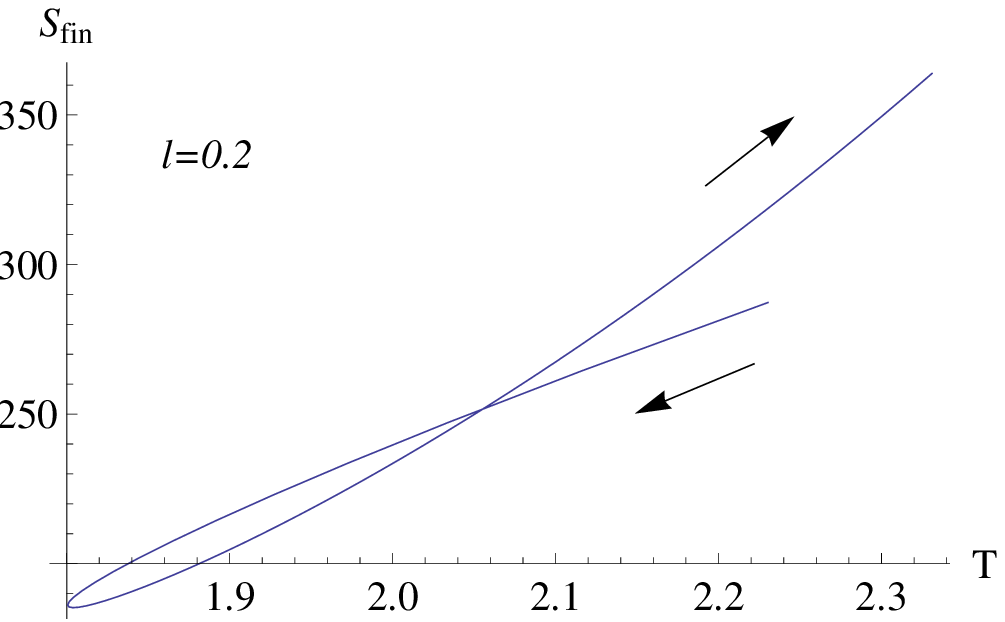}
        \includegraphics[scale=0.7]{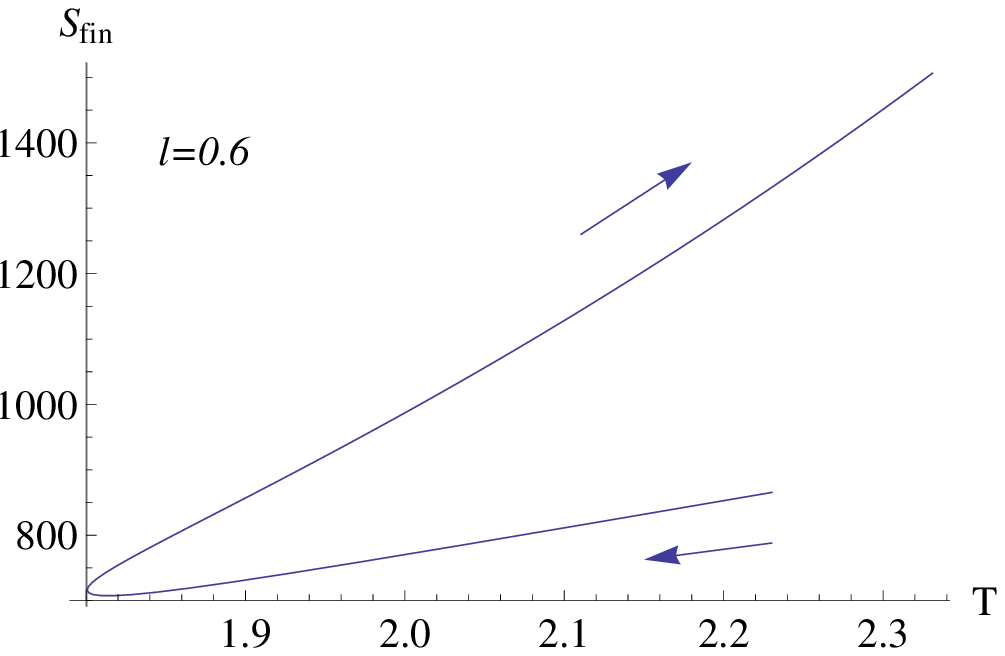}
        \includegraphics[scale=0.7]{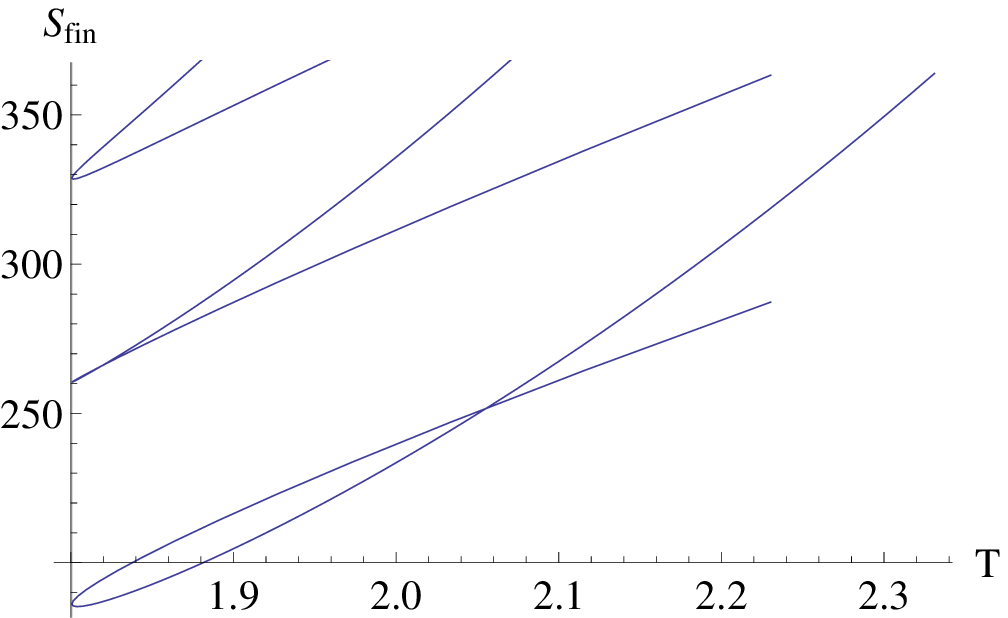}
        \includegraphics[scale=0.7]{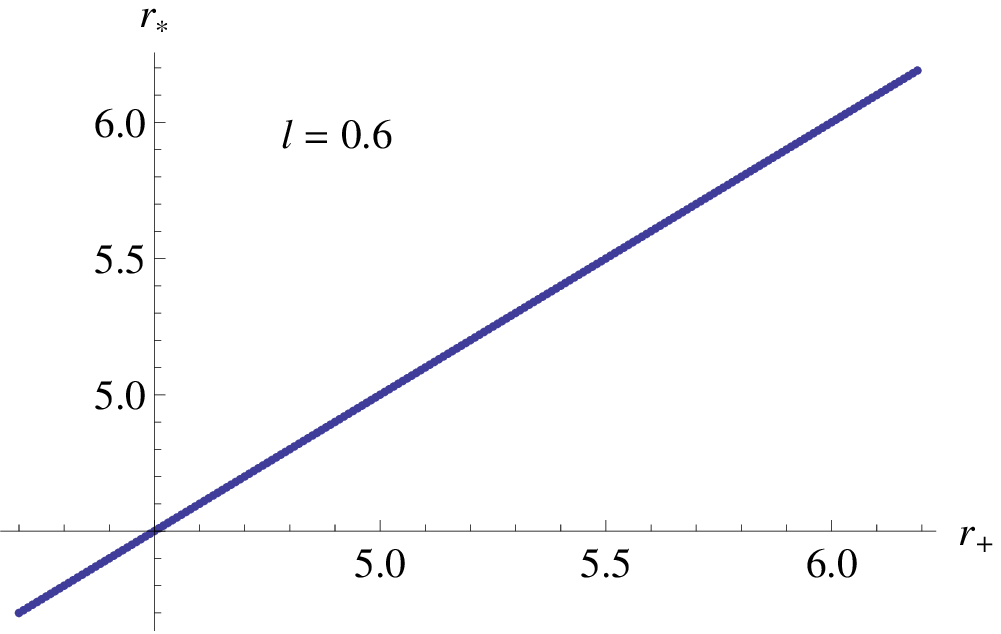}

        \caption{$S_{fin}(T)$ in the grand canonical ensemble with fixed $\phi=4$, and $l=0.2$ to $0.6$. The $r_*$ v.s. $r_+$ plot for $l=0.6$ showing the linear relation between the two as $l$ is large enough. The arrows indicate the direction of increasing $r_+$. Curves with $l=0.2$, $0.25$, $0.3$, from bottom to top are plotted together to exhibit the transition. }
        \label{Case1Grand}
\end{center}
\end{figure}

In the grand canonical ensemble, we take
\footnote{
In this paper, we choose the values $\phi=4$ and $Q=2$ just for convenience of performing numerical calculation. One can choose other values without any essential change of the results.
}
$\phi=4$ and consider the finite part of the entanglement entropy $S_{fin}$ as a function of the temperature. Figure \ref{Case1Grand} are the numerical results for some typical values of the strip width $l$. For $T>T_{min}$, there are two $S_{fin}$ values for one $T$, since there are two black hole solutions. At $T=T_{min}$, the specific heat $C_\phi$ diverges, implying a thermodynamical instability. The arrows around the plot denote the direction of increasing $r_+$.

For $l=0.2$, a distinctive feature is that the curve develops a self-intersection. In particular, starting from $T_{min}$ to the intersection point, the small black hole branch has a larger $S_{fin}$ compared with the large black hole branch, while the situation gets reversed beyond the intersection point. Note that this behavior is different from that of the thermal entropy $S_{th}(T)$ in Fig.  \ref{Case1Sthermal}, where the large black hole branch always has larger entropy. Of course, the entanglement entropy and thermal entropy are physically different by definition. In particular, via the AdS/CFT correspondence, they are given by different geometric quantities: the areas of the horizon and of the minimal surface. Thus they do not have to always share the similar behavior when $l$ is not large.

For larger value, $l=0.6$, the self-intersection behavior disappears and the plot becomes similar to the $S_{th}(T)$ plot. Indeed, for $l$ large enough, the minimal surface begins to wrap the horizon. This can also be inferred from the linear relation between $r_*$ and $r_+$ shown in Fig.  \ref{Case1Grand}. Therefore the thermal entropy begins to contribute dominantly to the finite part of the entanglement entropy. We also present the plots for $l=0.2$, $0.25$ and $0.3$ to show the emergence of the self-intersection.

\begin{figure}
\begin{center}
        \centering
        \includegraphics[scale=0.7]{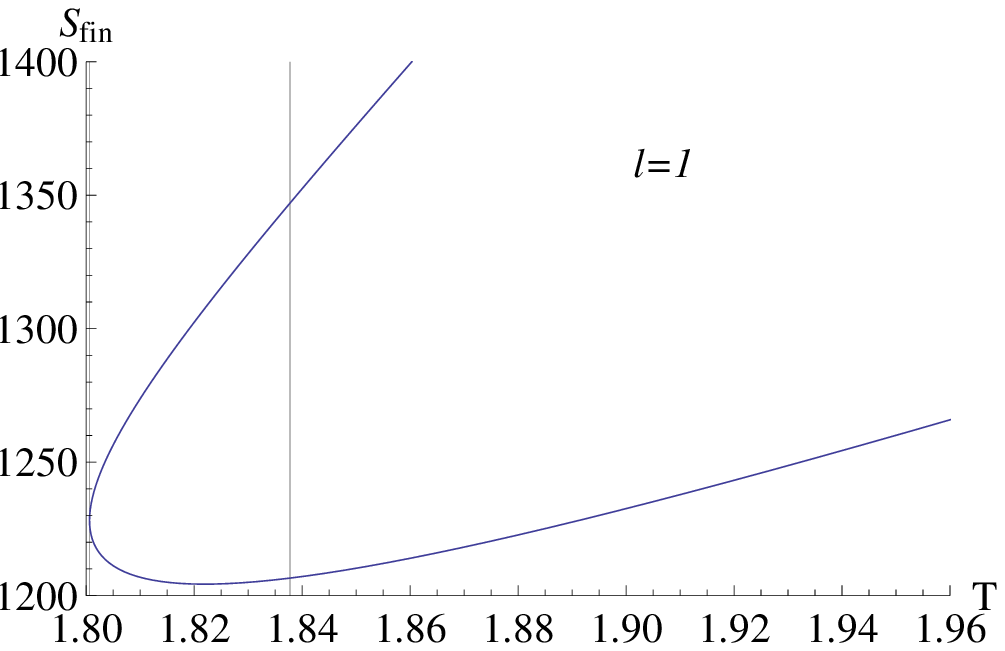}
        \includegraphics[scale=0.7]{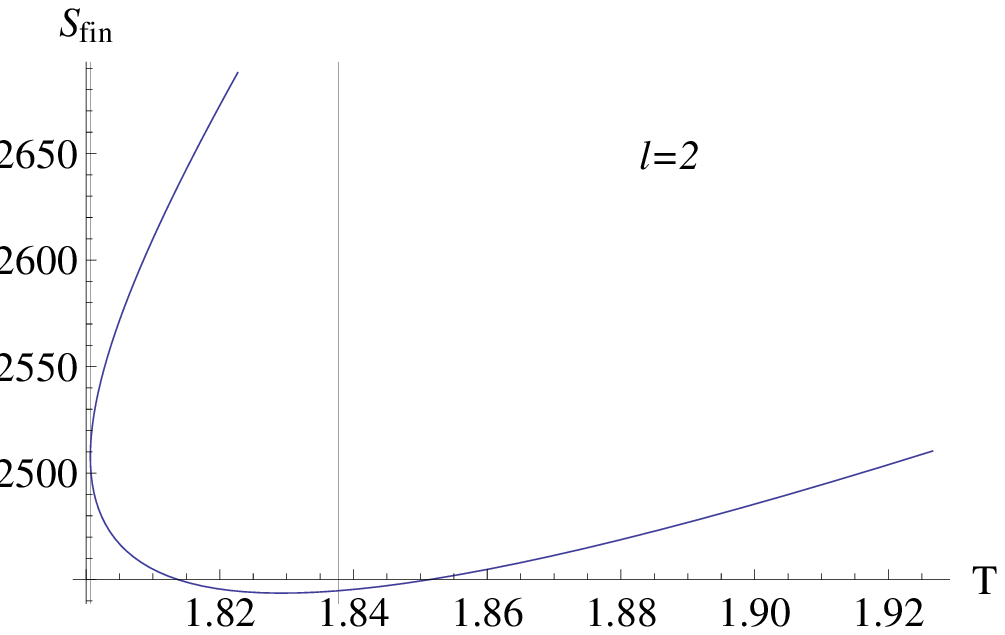}

        \caption{$S_{fin}(T)$ in the grand canonical ensemble with fixed $\phi=4$. The minimum moves from $T\approx1.82$ to $T\approx1.83$, as $l$ is increased from $1$ to 2. The vertical line to the right of $T=1.84$ is the position $T_1$ where the thermal entropy develops a minimum.}\label{Case1Zoom}
\end{center}
\end{figure}

Recall the thermal entropy $S_{th}(T)$ in Fig.  \ref{Case1Sthermal}, where the slope of the lower branch changes from negative to positive at a minimum $T_1\approx1.84$. We find similar behavior in the entanglement entropy when $l$ is large. In particular, as the value of $l$ is increased, this minimum moves towards $T_1$ from the left. In Fig.  \ref{Case1Zoom}, one can see that the position of the minimum moves from  $T\approx1.82$ to $T\approx1.83$, as $l$ is increased from $1$ to 2. It is expected that as $l$ gets larger, the holographic entanglement entropy will asymptotically recover the thermal entropy.

\begin{figure}
\begin{center}
        \centering
        \includegraphics[scale=0.7]{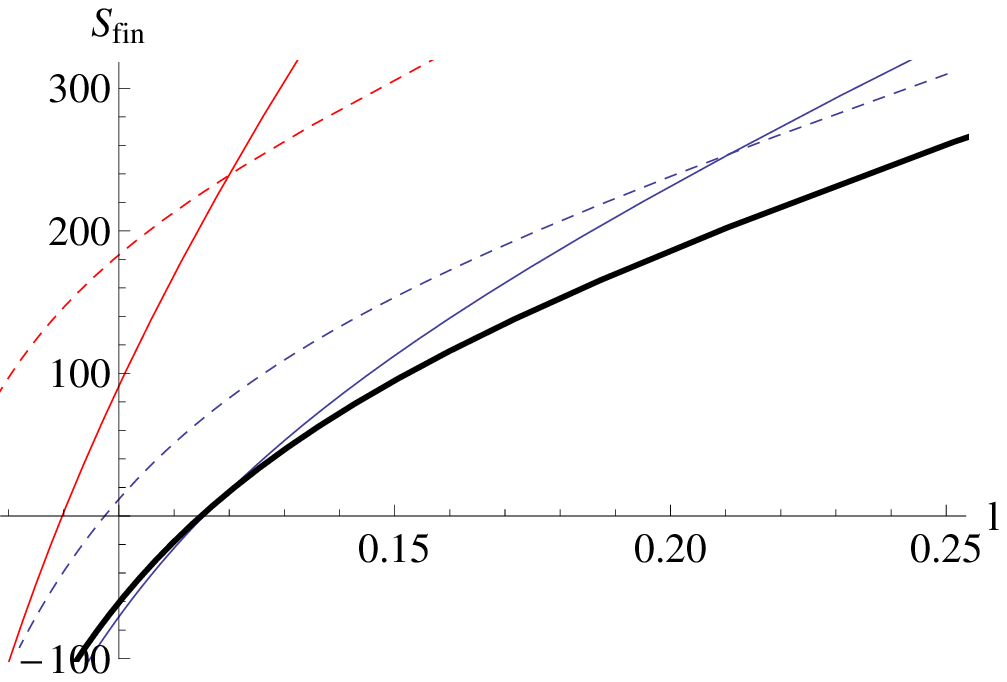}

        \caption{$S_{fin}(l)$  in the grand canonical ensemble with fixed $\phi=4$,  $T=73/(\sqrt{65}\pi)\approx2.88$ (red), $T=14/(\sqrt{5}\pi)\approx1.99$ (blue). The solid curves correspond to the large black hole branches, and the dashed curves to the small black hole branches. The black solid curve corresponds to $T_{min}\approx1.80$, where the two branches merge into one. }
        \label{Case1GrandSL}
\end{center}
\end{figure}

 One can also study $S_{fin}$ as a function of the strip width $l$ for fixed temperatures. In Fig.  \ref{Case1GrandSL}, we plot $S_{fin}(l)$ for two different temperatures $T=14/(\sqrt{5}\pi)\approx1.99$ (blue) and $T=73/(\sqrt{65}\pi)\approx2.88$ (red). The solid curves correspond to the large black hole branches while the dashed ones to the small black hole branches. From these plots, one can see that $S_{fin}$ generically increases with $T$. Moreover, at a fixed $T$, starting from small $l$, the large black hole branch has smaller $S_{fin}$. When $l$ increases to certain value, the two branches have the same $S_{fin}$, and after that the small black hole branch has smaller $S_{fin}$. Such cross-over behavior corresponds to the self-intersection of the $S_{fin}(T)$ plot. As the temperature is lowered, the open angle of the two branches of the same temperature gets smaller. At $T_{min}$, the two branches coincide, as represented by the thick black curve in the figure. Note that here the multiple value behavior of $S_{fin}(l)$ is due to the existence of {\it two branches} of solutions (geometries) at a fixed temperature. In contrast, in the study of the  confinement/deconfinement-type phase transitions  of \cite{Nishioka:2006gr,Klebanov:2007ws,Faraggi:2007fu}, the multiple value behavior there arises from the existence of two possible minimal surfaces in {\it one fixed} geometry.

Note also that for small $l$, the minimal surface becomes localized in the asymptotic AdS boundary, thus $S_{fin}(l)$ should recover the result in pure AdS$_5$ \cite{Ryu:2006ef}
\be
S_A^{AdS}(l)=\frac{1}{4G}\left[ V_2r_c^2-4\pi^{3/2} \left(\frac{\Gamma(\frac{2}{3})}{\Gamma(\frac{1}{6})}\right)^3 \frac{V_2}{l^2}\right]=S_{div}+S_{fin}^{AdS}(l),
\ee
where the divergent part $S_{div}\propto V_2$ gives the well-known `area law' \cite{Bombelli:1986rw,Srednicki:1993im}, and $S_{fin}^{AdS}(l)\sim -2.0148/l^2$, indicating a power law decay of the magnitude of the finite part with increasing $l$ . Indeed, our results  exhibit such power law behavior for small $l$. For example, when $l\lesssim0.014$, our $S_{fin}(l)$ for fixed $r_+=6$ (i.e. the large black hole branch at $T=14/(\sqrt{5}\pi)$) can be fitted by $S_{fin}(l)\sim -2.0/l^2$, which is consistent with $S_{fin}^{AdS}(l)$.
However, pure AdS provides no reference for the behavior of $S_{fin}(T)$ in our case when the minimal surface is localized at the asymptotic boundary for small $l$.  Indeed, in pure AdS with the period of the Euclidean time corresponding to the temperature $T$, the minimal surface in the bulk is not affected by variations of $T$.

\begin{figure}
\begin{center}
        \centering
        \includegraphics[scale=0.7]{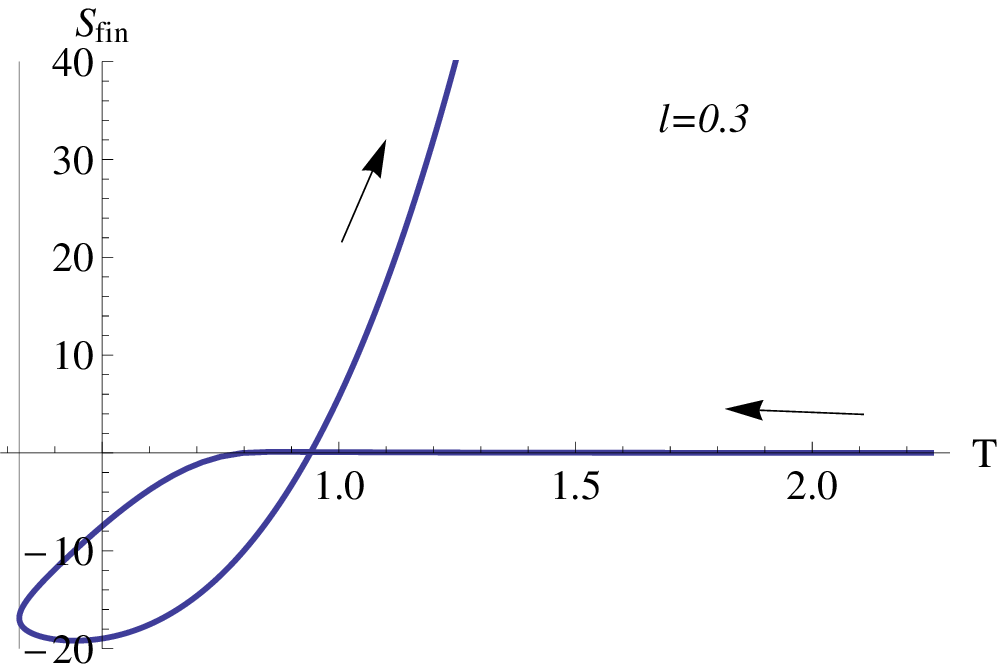}
        \includegraphics[scale=0.7]{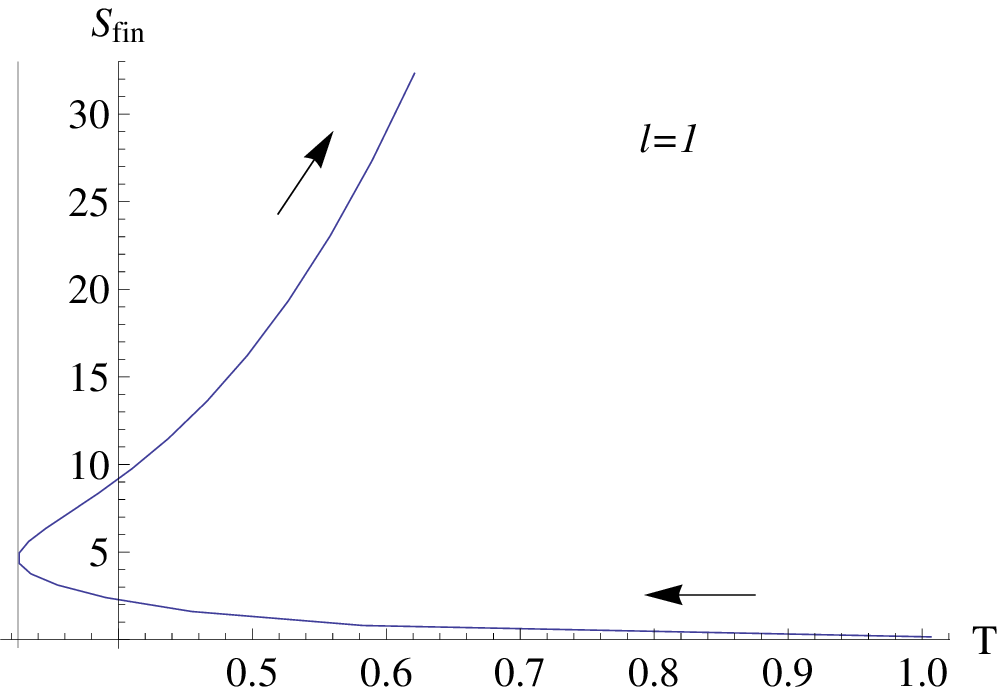}

        \caption{$S_{fin}(T)$ in the canonical ensemble with fixed $Q=2$, $l=0.3$ and $1$. The arrows indicate the direction of increasing $r_+$. The vertical line denotes the temperature at which the specific heat diverges.} \label{Case1Cano}

\end{center}
\end{figure}

\begin{figure}
\begin{center}
        \centering
        \includegraphics[scale=0.7]{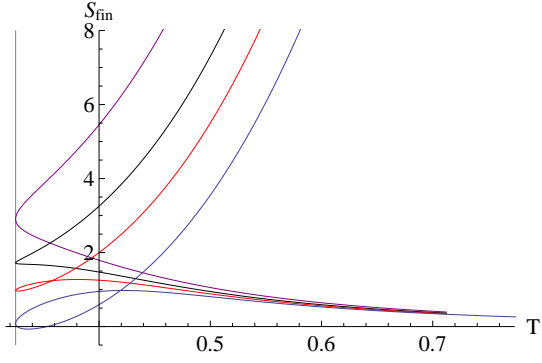}

        \caption{$S_{fin}(l)$ in the canonical ensemble with fixed $Q=2$, and, from top to bottom, $l=0.8,\ 0.7,\ 0.65,\ 0.6$ } \label{Case1CanoSTtransition}

\end{center}
\end{figure}

In the canonical ensemble with fixed $Q=2$, the results are shown in Fig.  \ref{Case1Cano}. For $l=0.3$, starting from $ T_{min}\approx0.32$, the small black hole (upper) branch increases until around $T=0.8$, then gradually decreases to zero. The large black hole branch (lower) starting from $T_{min}$ first decreases a little bit and then begins to increase after around $T=0.5$. The two branches intersect at about $T=0.9$. For $l=1$, such self-intersection disappears and the plot resembles the thermal entropy. Note that in this case, the small black hole branch always has a negative specific heat. The emergence of the self-intersection is exhibited in Fig.  \ref{Case1CanoSTtransition} for the plots with $l$ between $0.6$ and $0.8$.


\begin{figure}
\begin{center}
        \centering
        \includegraphics[scale=0.7]{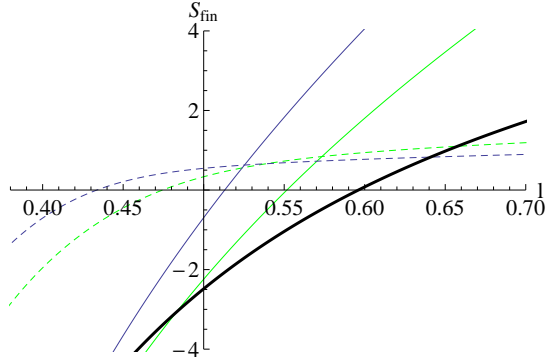}

        \caption{$S_{fin}(l)$ in the canonical ensemble with fixed $Q=2$,  $T=0.51$ (blue), $T=0.45$ (green). The solid curves correspond to the large black hole branches, and the dashed curves to the small black hole branches. The black solid curve corresponds to $T_{min}\approx0.32$, where the two branches merge into one. } \label{Case1CanoSL}

        \end{center}
\end{figure}

The function $S_{fin}(l)$ in the canonical ensemble is plotted in Fig.  \ref{Case1CanoSL} for $T=0.51$ (blue) and $T=0.45$ (green). The thick black curve corresponds to the minimal temperature $T_{min}\approx0.32$. Again, as $T$ gets closer to $T_{min}$, the open angle of the two branches becomes smaller. Finally the two branches merge into the black curve at $T_{min}$. Note, in contrast to the result of the grand canonical ensemble in Fig. \ref{Case1GrandSL}, the dashed curves here intersect with each other, corresponding to the behavior of $S_{fin}(T)$ in Fig. \ref{Case1Cano} and \ref{Case1CanoSTtransition} that $S_{fin}$ of the small black hole branch decreases as $T$ (after the self-intersection point, if any) increases.

\subsection{Single-Charge Black Holes in AdS with $D=4$ and $7$}

In general, the metric for these R-charged black holes can be summarized as
\be
ds^2_{D}=-\ch^{b-1}fdt^2+\ch^b\left[ \frac{dr^2}{f}+r^2(dx^2+d\vec y^2_{D-3})\right] ,
\ee
where $\ch \equiv \Pi_i^n H_i(r)$, $n=4,3,2$ for $D=4,5,7$, respectively, $b=\frac{1}{D-2}$, and
\be
H_i=1+\frac{q_i}{r^{D-3}},\ \ \  f=-\frac{\mu}{r^{D-3}}+\frac{r^2}{L^2}\ch.
\ee
The temperature of the black holes is given by
\be
T=\frac{f'(r_+)}{4\pi\sqrt{\ch(r_+)}},
\ee
where $r_+$ again denotes the horizon at which $f(r_+)=0$. The thermal entropy is
\be
S_{th}=\frac{A}{4G}=2\pi \sqrt{\ch(r_+)}r_+^{D-2}.
\ee
The details of the thermodynamics can be found in \cite{Cvetic:1999ne,Cvetic:1999rb,Sahay:2010yq}, and they are almost similar to the results in $D=5$ (except for the canonical ensemble in $D=4$, see below), so will not be listed here.

The general form of the holographic entanglement entropy in the strip configuration is
\ba
l(r_*)&=&2\int^\infty_{r_*}\frac{dr}{\sqrt{\left(\frac{\ch}{\ch_*}\frac{r^{2(D-1)}}{r_*^{2(D-2)}}-r^2\right)f}},
\\
S_A(r_*)&=&\frac{V_{D-3}}{2G}\int^{r_c}_{r_*} r^{D-3}\sqrt{\ch}\sqrt{\frac{\ch}{\ch_*}\frac{r^{2(D-1)}}{r_*^{2(D-2)}}}\frac{dr}{\sqrt{\left(\frac{\ch}{\ch_*}\frac{r^{2(D-1)}}{r_*^{2(D-2)}}-r^2\right)f}}.
\ea
We will still set $V_{D-3}=1$ for convenience. The UV  regularized expression can be obtained from
\be
S_A(r_*)=\frac{r_c^{D-3}}{2(D-3)G}+S_{fin}(r_*).
\ee

\begin{figure}
\begin{center}
        \centering
        \includegraphics[scale=0.7]{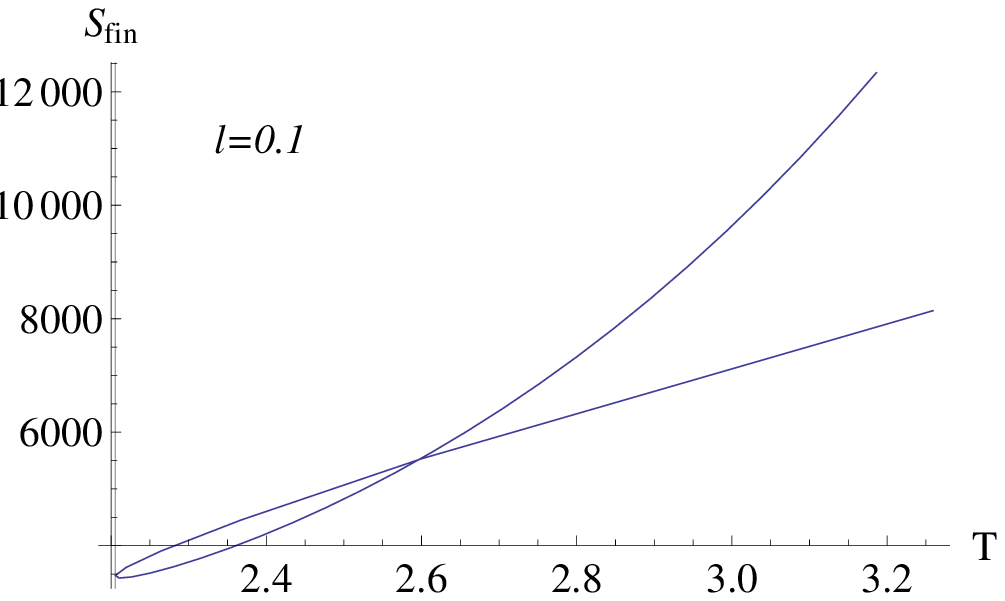}
        \includegraphics[scale=0.7]{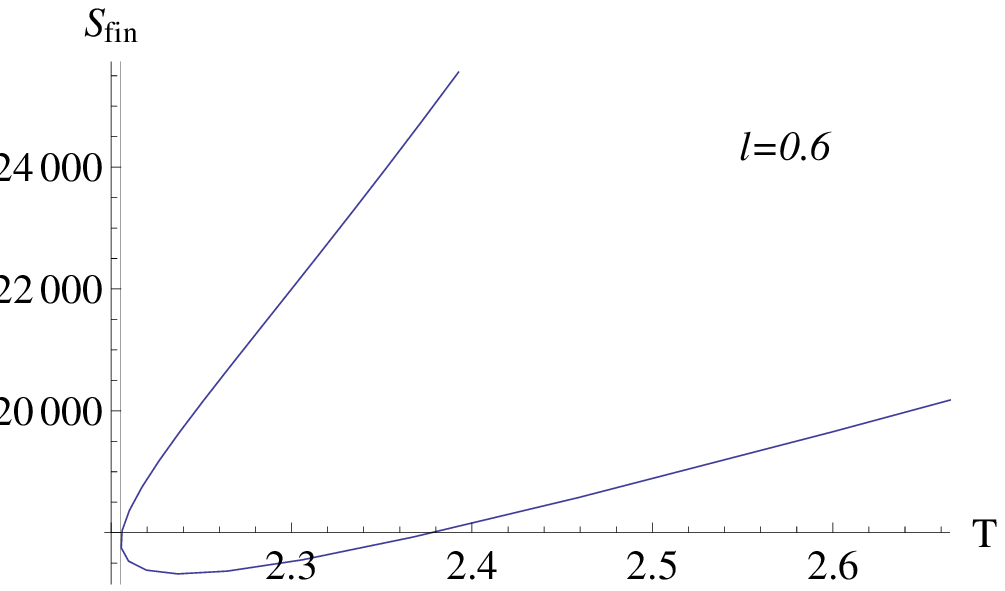}

        \caption{$S_{fin}(T)$ in AdS$_7$ in the grand canonical ensemble with fixed $\phi=4$, $l=0.1$ and $0.6$.} \label{AdS7Case1Grand}

        \end{center}
\end{figure}

\begin{figure}
\begin{center}
        \centering
        \includegraphics[scale=0.7]{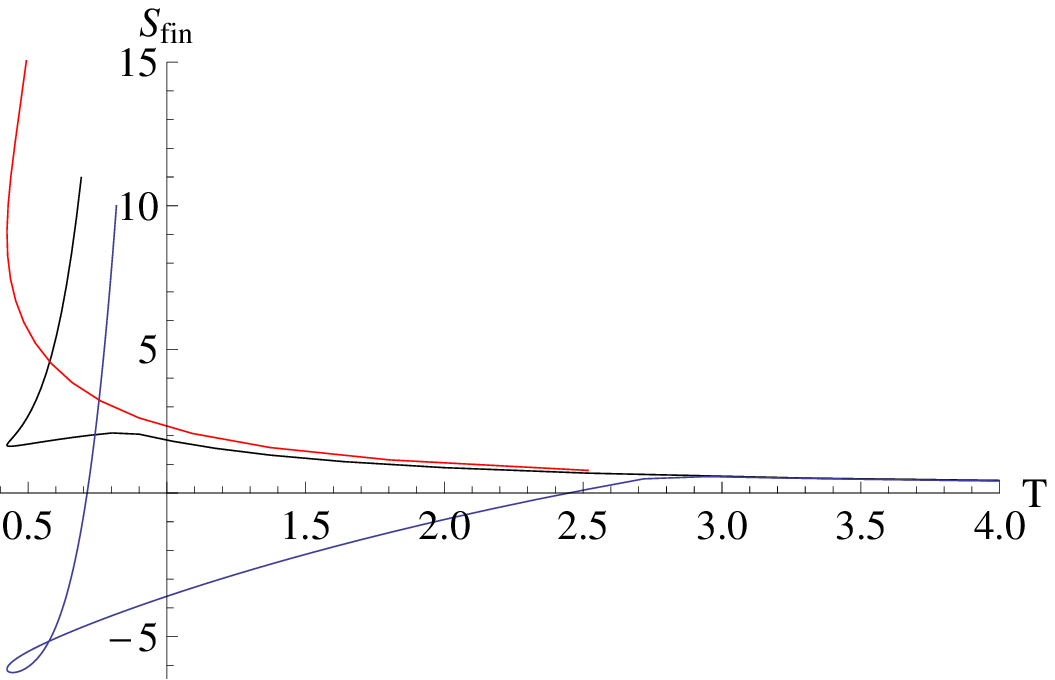} 
        \includegraphics[scale=0.7]{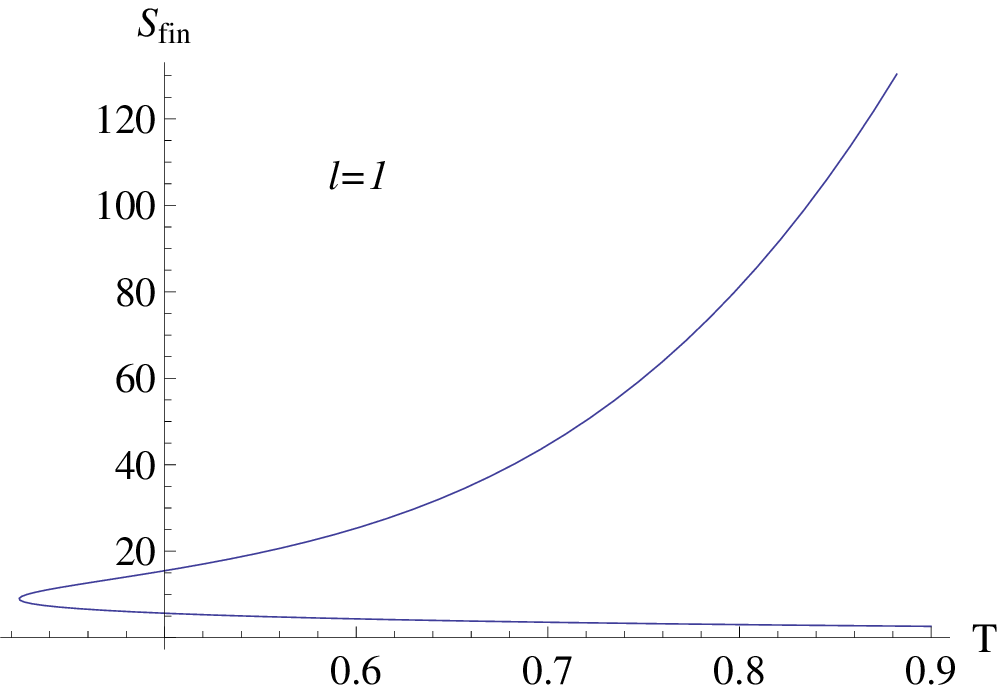}

        \caption{$S_{fin}(T)$ in AdS$_7$ in the canonical ensemble with fixed $Q=2$. Left: transition from $l=0.3$ (bottom, blue), $l=0.4$ (middle, black) and $l=1$ (top, red). Right: $l=1$. } \label{AdS7Case1Cano}
\end{center}
\end{figure}

\begin{figure}
\begin{center}
        \centering
        \includegraphics[scale=0.7]{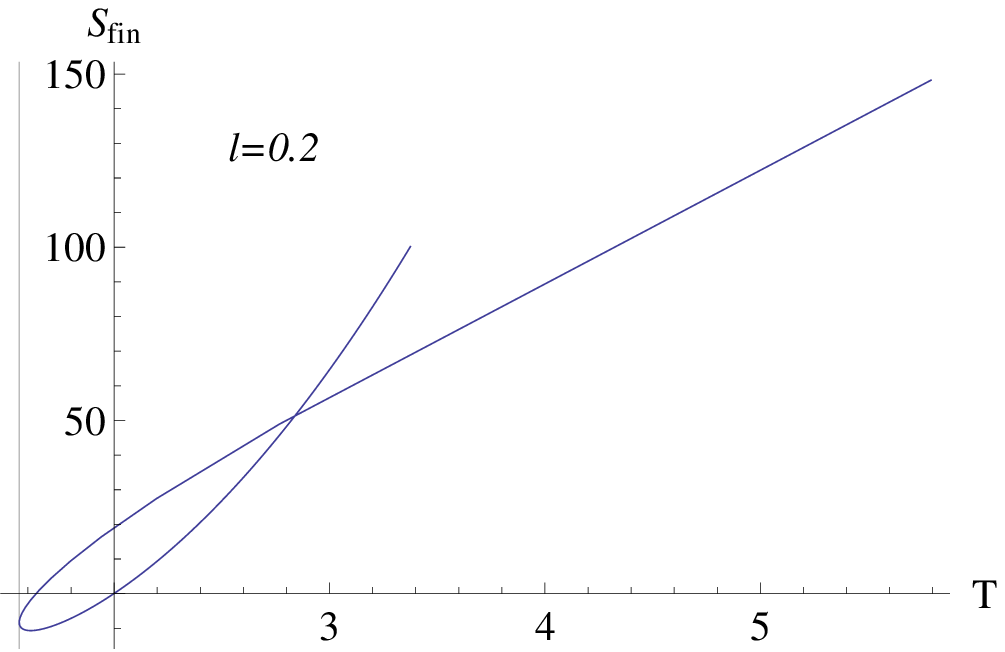}
        \includegraphics[scale=0.7]{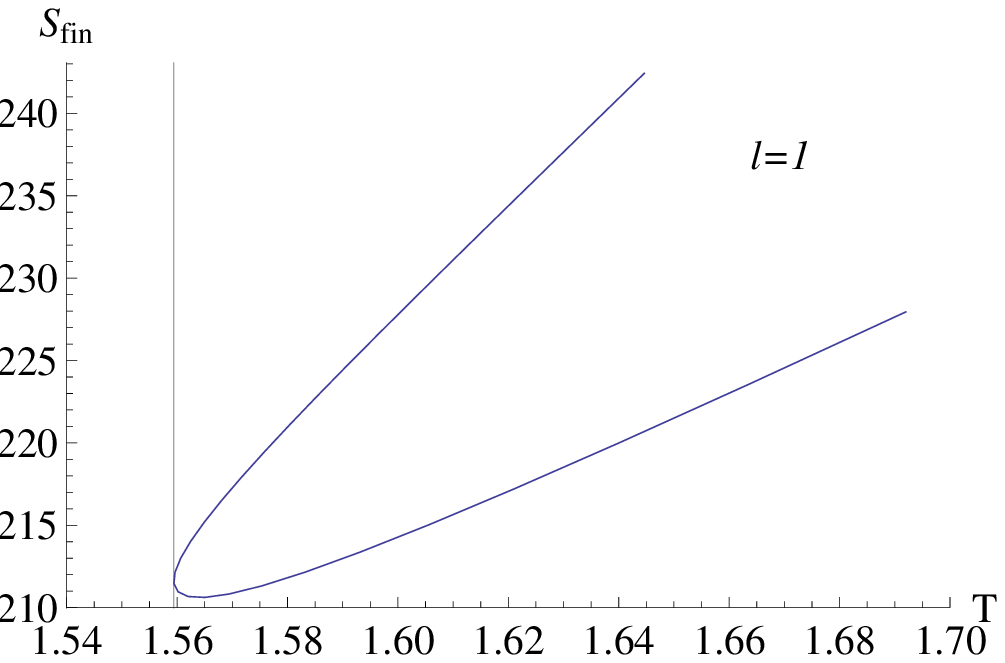}

        \caption{$S_{fin}(T)$ in AdS$_4$ in the grand canonical ensemble with fixed $\phi=4$, $l=0.2$ and $0.6$.}
        \label{AdS4Case1Grand}

         \end{center}
\end{figure}

We find that most of the results in  $D=4$ and $7$ are similar to those of  $D=5$. So we will be brief here and only present their typical plots. In the grand canonical ensemble in $D=4$ and $7$, the $S_{fin}(T)$ plot in Fig.  \ref{AdS7Case1Grand} and \ref{AdS4Case1Grand} develops a self-interaction for small $l$, while it resembles the thermal entropy for large $l$. In Fig.  \ref{AdS7Case1Cano}, the canonical ensemble in $D=7$ exhibits similar behavior, which is essentially analogous to the case of $D=5$.

\begin{figure}
\begin{center}
        \centering
        \includegraphics[scale=0.7]{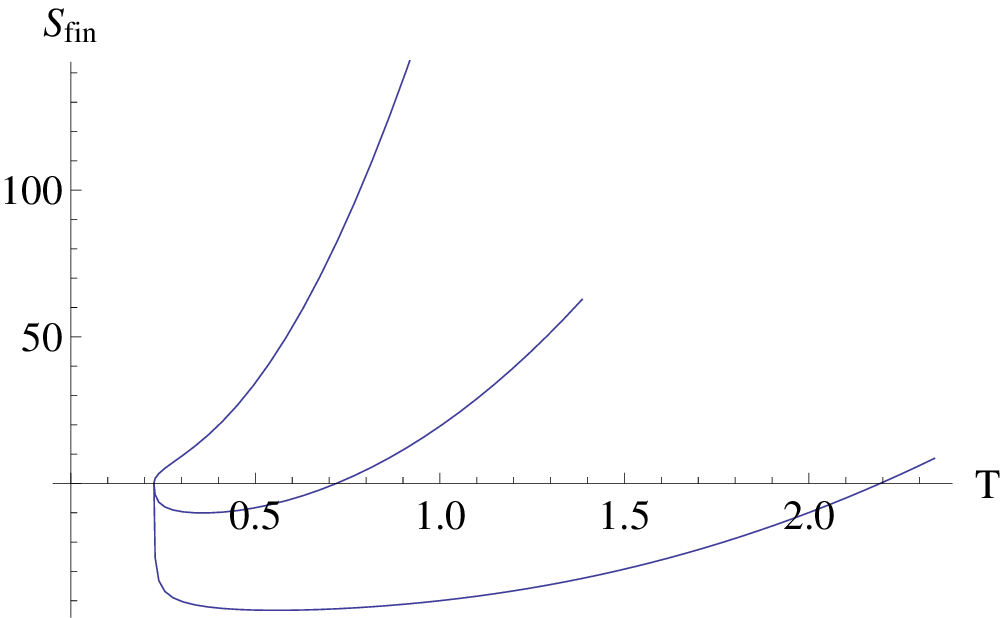}
         \includegraphics[scale=0.7]{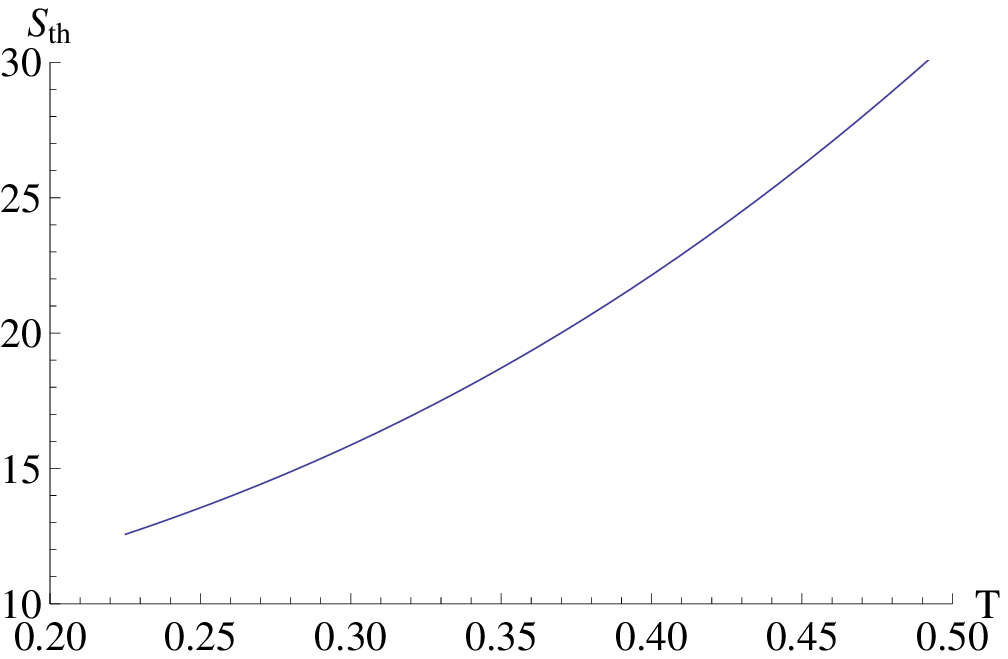}

        \caption{$S_{fin}(T)$ in AdS$_4$ in the canonical ensemble with fixed $Q=2$. Left: $S_{fin}(T)$ plots with $l=2$, $0.6$ and $0.2$, from top to bottom. Right: $S_{th}(T)$, starting from the minimum $T_{min}=\frac{\sqrt{Q}}{2\pi}\approx0.225079$, $S_{th}\approx12.56$. }\label{AdS4Case1Cano}
\end{center}
\end{figure}

\begin{figure}
\begin{center}
        \centering
        \includegraphics[scale=0.7]{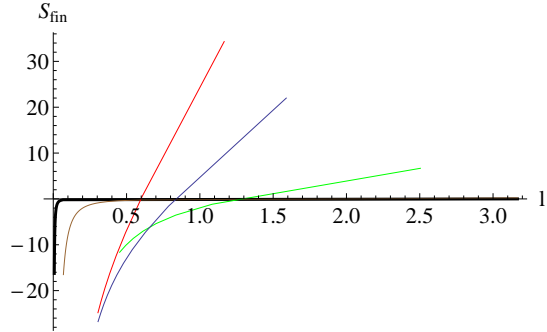}

        \caption{$S_{fin}(l)$ in AdS$_4$ in the canonical ensemble with fixed $Q=2$, and, from top to bottom, $T=0.72$ (red), $0.49$ (blue), $0.25$ (green), $0.22509$ (brown) and $T_{min}$ (thick black).  }\label{AdS4ExceptionSL}
\end{center}
\end{figure}

An exception is the case of $D=4$ in the canonical ensemble, where  the single-charge black holes in AdS$_4$ (or the M2-branes with a single angular momentum) have no thermal instability \cite{Cai:1998ji, Cvetic:1999rb}, as can be seen from the monotonic behavior of $S_{th}(T)$ in the right plot of Fig.  \ref{AdS4Case1Cano}. In this case, the relation $S_{th}(T)$ is determined by
\be
S_{th}(r_+)=\pi(r_+^2+\sqrt{4Q^2+4r_+^4}),\ \ \ T(r_+)=\frac{2r_+^2+\sqrt{4Q^2+r_+^4}}{2\pi\sqrt{2r_+^2+2\sqrt{4Q^2+r_+^4}}}.
\ee
Note that at $r_+=0$, $T$ has a minimum $T=\sqrt{Q}/(2\pi)$  and correspondingly $S_{th}=2\pi Q$.
The left plot in Fig.  \ref{AdS4Case1Cano} exhibits $S_{fin}(T)$ with different $l$. In particular, for large $l$, i.e. $l=2$, $S_{fin} $ monotonically increases with $T$, whereas for small $l$, i.e. $l=0.2$ and $0.6$, $S_{fin}$ starting from the minimal temperature first decreases to some minimum, then becomes an increasing function of $T$. The corresponding $S_{fin}(l)$ is given in Fig. \ref{AdS4ExceptionSL} with $Q=2$, from which one can see that the slope of the curves decreases  as the temperature is lowered. In particular, it is not hard to infer that as $T$ becomes infinitesimally close to $T_{min}$, represented by the thick black curve, \footnote{ Due to numerical difficulties, we cannot draw the plot exactly at $T_{min}$, where $r_+=0$. But we can approximate it by the thick black curve with $r_+=10^{-4}$ and $T\approx0.225079$ very close to $T_{min}=\frac{1}{\sqrt{2}\pi}$. } the neighboring curve intersects the thick black curve at some definite value, denoted by $l_c$ for later convenience.

\section{Discussion}

\begin{figure}
\begin{center}
        \centering
        \includegraphics[scale=0.7]{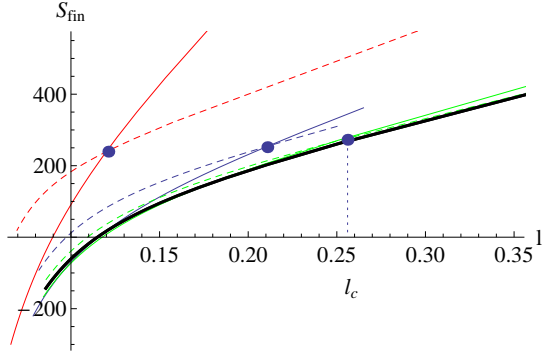}

        \caption{Illustration for the critical value $l_c$. $T=$ $2.88$ (red), $1.99$ (blue), $1.83$ (green) and $T_{min}=1.80$ (black). Solid curves correspond to large black hole branch and dashed ones to small black hole branch. The dots denote the intersection points. For a fixed $l\geq l_c$, $S_{fin}$ for the small black holes is always lower than that of the large black holes. }\label{Case1Canolc}
\end{center}
\end{figure}

For the R-charged black holes in AdS with $D=4,5,7$, we find that the holographic entanglement entropy can reflect the information about thermodynamic instability. In particular, in the case of a single charge, the derivative  $S_{fin}'(T)$ diverges at $T_{min}$, indicating the onset of a thermodynamic  instability. This coincides  with the behavior of the thermal entropy $S_{th}'(T)$.

Compared to the dependence of the thermal entropy on the temperature, there is an extra parameter, i.e. the width of the subsystem $l$, that controls the behavior of the entanglement entropy $S_{fin}(T)$. As we have seen above, as $l$ increases, the behavior of the entanglement entropy becomes more similar to that of the thermal entropy, which is expected since with larger $l$, the minimal surface droops deeper into the IR and begins to wrap the horizon, therefore the thermal contribution dominates in the holographic entanglement entropy. For small $l$, on the other hand, the minimal surface is localized around the asymptotic boundary. Therefore the two types of entropies may not necessarily resemble each other. In the case of RNAdS black holes with a finite volume discussed in \cite{Johnson:2013dka}, the entanglement entropy resembles the thermal entropy even when $l$ is not large (c.f. also \cite{Cai:2012nm} in the case of a holographic p-wave superconductor model). As noted in \cite{Johnson:2013dka}, however, the entanglement entropy plot for small $l$ dose not exactly coincide with $S_{th}(T)$, even after an overall scaling.  In contrast, our results indicate that for the R-charged black holes, $S_{fin}(T)$ is significantly different from its thermal cousin for small $l$. In particular, in the low temperature region, the $S_{fin}(T)$ plot develops an self-intersection.

The qualitative change of the plot as one varies $l$ seems to indicate some sort of phase transition characterized by a critical value $l_c$ of the width of the strip. This can be illustrated more clearly in the $S_{fin}(l)$ plot. Consider the  case of the single-charge black holes in the grand canonical ensemble as an example; other cases are similar
\footnote{In particular, the similar argument applies to the case of $D=4$ in the canonical ensemble, where $l_c$ characterizes the qualitative change of the $S_{fin}(T)$ curves between the monotonic ones for $l\geq l_c$ and the non-monotonic ones for $l<l_c$.   The existence of a particular value $l_c$ can be illustrated by Fig. \ref{AdS4ExceptionSL}.  }.
As one can see from Fig.  \ref{Case1Canolc}, when $T$ is lowered all the way to $T_{min}$, the intersection point of the two branches at the same temperature approaches a certain value $l_c$ around $0.25$. If one fixes the width to some value $l\geq l_c$, then the small  black hole branch (dashed curves) will always have smaller $S_{fin}$ for all $T\geq T_{min}$, indicating the absence of any self-intersection in the $S_{fin}(T)$ plot. This is analogous to the case of the confining geometries studied in \cite{Nishioka:2006gr,Klebanov:2007ws,Faraggi:2007fu}, where there also exists a critical value for the width of the strip governing the confinement/deconfinement-type phase transition. As noted above, the critical value there characterizes the transition between a connected and a disconnected extremal surfaces of the same geometry. On the other hand, the critical value studied here reflects the qualitative change of the contributions from the two branches of small and large black hole solutions.
It would be interesting to further investigate what this corresponds to in the dual CFT.

Moreover, although the precise meaning of such a scale $l_c$ is not quite clear so far, at least, its existence seems to be related to the fact that the strip subsystem on the boundary breaks the rotational invariance. Indeed, the scale introduced by this anisotropy is expected to play an essential role in the phase transition of a strip subsystem in the background of extremal charged dilatonic black holes \cite{Kulaxizi:2012gy,Erdmenger:2013rca}, where it was shown that the phase transition is absent when the subsystem is a sphere (which preserves the isotropy), and that an annulus maybe regarded as a configuration interpolating between the strip and the sphere as two limiting cases. It would be more illuminating to further study other subsystems such as a sphere or an annulus for comparison with our results obtained using a strip.

In fact, one can also study the case  with two equal charges, i.e. $q_1=q_2=q$, $q_3=0$. In this  (non-extremal
\footnote{
 The holographic entanglement entropy in the extremal case has been studied in \cite{Alishahiha:2012ad,Kulaxizi:2012gy}.
})
case, it can be easily checked that the thermodynamics is somewhat trivial; e.g. the thermal entropy is a monotonic function of the temperature. However, there is a thermodynamic instability associated with the divergence of a properly defined susceptibility \cite{Gubser:2009qt} when the two charges are equal. Such instability is not reflected in the specific heat. Indeed, simple calculation indicates that the specific heat is always positive. In other words, the thermal entropy carries no information of this instability since its derivative with respect to the temperature is always positive. We have checked that, similar to $S_{th}$, the entanglement entropy cannot probe this thermodynamic instability, either.

Our work in planar R-charged black holes with $k=0$ can be extended to the case of $k=1$ with a spherical horizon. The thermodynamical analysis in \cite{Cvetic:1999ne,Cvetic:1999rb,Sahay:2010yq} indicates that this case may have richer phase structures than the planar case, since the latter can be regarded as the large black hole limit of the former. Of course, the subsystem should be changed to some configurations more appropriate for the spherical case, e.g. a disk or an `orange slice' as discussed in \cite{Johnson:2013dka}. We leave this study for future work.

\section*{Acknowledgements }
The author would like to thank Rong-Gen Cai for indispensable suggestions and invaluable discussions. Special thanks also go to Song He and Li Li for helpful discussions. This work was supported in part by the project of Knowledge Innovation Program of Chinese
Academy of Science, NSFC under Grant No.11175225,
and National Basic Research Program of China under
Grant No.2010CB832805.

\bibliographystyle{apsrev}
\bibliography{Biblio}

\end{document}